\documentclass[aps,pra,floatfix,twocolumn,showpacs,nofootinbib]{revtex4-1}
\usepackage[english]{babel}

\usepackage{graphicx,times}                    
\usepackage{amsmath,amssymb,amsthm}        
\usepackage{subfigure}
\newcommand{\ket}[1]{\left | #1 \right \rangle}	
\newcommand{\bra}[1]{\left \langle #1 \right |}		
\newcommand{\proj}[2]{\left | #1 \right \rangle \! \left \langle #2 \right |}	
\newcommand{\hta}{{a}^\dagger}		
\newcommand{\I}{\imath}
\newcommand{\abs}[1]{\left | #1 \right |}		
\newcommand{\abss}[1]{\left | #1 \right |^2}		
\newcommand{\tr}[1]{ \mathtt{Tr} \! \left [ #1 \right ] }		
\newcommand{\Tr}[2]{\mathtt{Tr}\!_{#1}\! \left [ #2 \right ] }		

\begin{document}
\title{Experimental quantum tomography of a homodyne detector}
\author{Samuele Grandi}\email{s.grandi@imperial.ac.uk}
\affiliation{Centre for Cold Matter, Blackett Laboratory, 
Imperial College London, Prince Consort Road, SW7 2AZ London, UK}
\author{Alessandro Zavatta}
\email{alessandro.zavatta@ino.it}
\author{Marco Bellini}
\email{marco.bellini@ino.it}
\affiliation{Istituto Nazionale di Ottica (INO-CNR), 
Largo Enrico Fermi 6, I-50125 Florence, Italy}
\affiliation{LENS and Department of Physics and Astronomy, 
University of Florence, I-50019 Sesto Fiorentino, Florence, Italy}
\author{Matteo G. A. Paris}
\email{matteo.paris@fisica.unimi.it}
\homepage{users.unimi.it/aqm}
\affiliation{Quantum Technology Lab, 
Dipartimento di Fisica, Universit\`a degli 
Studi di Milano, I-20133 Milano, Italy}
\affiliation{CNISM, UdR Milano Statale, I-20133 Milano, Italy}
\date{\today}
\begin{abstract}
We suggest and demonstrate a tomographic method to fully characterize
homodyne detectors at the quantum level. The operator measure associated
with the detector is expanded in the quadrature basis and probed with a
set of coherent states. The coefficients of the expansion are 
retrieved using a least squares algorithm. We then validate 
the reconstructed operator measure on nonclassical states. 
Finally, we exploit results to estimate 
the overall quantum efficiency of the detector.
\end{abstract}
\maketitle
\section{Introduction}
Balanced homodyne detection is a crucial detection technique for
continuous variable quantum technology and lies at the core of many
quantum optics experiments \cite{Lvovsky2009,Dariano2003,dgw99}. 
From its first proposal for the measurement
of quadrature squeezing, to its current extensive use in the fields of 
quantum tomography, quantum communication and quantum metrology 
\cite{Vogel1989,Smithey1993,Dariano1994,ad02,por09,wit10,lvo11,hodi12,kumar12,
hcom13,Berni15,sb16},
this detection scheme has carved its place
into experimental quantum optics. Besides quantum optical systems, homodyne detection extends
its reach to the whole field of continuous variable quantum
technologies, spanning from atomic
systems \cite{ell07,Gross2011} to quantum optomechanics \cite{Verhagen2012}. 
\par
Advances in technology promoted the spread of many different
configurations of this versatile apparatus, tailored to disparate
experimental needs. Such a wide range of applications calls for reliable
ways to fully characterize homodyne detectors. Each specific setup
relies on classical calibrations in order to gain the most general
description of the apparatus and of the relationship between the input
state and the measurement output. Recently, a characterisation of
homodyne detection, only used as a phase-insensitive photon counter, was
demonstrated \cite{Cooper2014}. However, a general and reliable model
for the description of the fully phase-sensitive homodyne 
detection in the form of a Quantum Detector Tomography (QDT) is, 
in fact, still lacking.
\par
The pioneering proposals for QDT 
\cite{Luis1999, Fiurasek2001, Dariano2004,
Dariano2007} were followed by the experimental
characterization of an avalanche photodiode, in both single and
time-multiplexed configurations, for the detection of up to eight
photons \cite{Lundeen2009}. Subsequent works developed the idea,
including the effect of decoherence onto the operator description
\cite{Dauria2011}, or different detection devices, such as
superconducting nanowires \cite{Akhlaghi2011} and TES based systems
\cite{Brida2012, Brida2012a}. However, these are detectors devoted to
photon counting, whose description is entirely embedded in the 
diagonal sector of the Fock
space. Only recently, a specific phase-sensitive hybrid scheme, in the 
form of a weak homodyne detector based on photon counting, was the 
object of an experimentally realized QDT \cite{Zhang2012}. 
In this paper, we move several crucial steps forward, and present
a theoretical and experimental realization of QDT for 
homodyne detector, i.e.  the most
commonly used form of a fully phase-sensitive detector, whose operators
are naturally described in phase space.
\par 
The quantum description of any detector is given by a
positive operator-valued measure (POVM), i.e. a set of positive
operators $\left \{ \Pi_n \right \}$, 
giving a resolution of identity $\sum_n \Pi_n = {\mathbb I}$. 
The determination of these operators is, in turn, the main goal of
detector tomography. Given an input state $\rho$, the Born rule states
that $ p_n^\rho = \tr{\rho \, \Pi_n}$ is the probability of obtaining
the outcome $n$ when the generalized observable described by $\Pi$ 
is being measured. The inversion of this
formula allows the reconstruction of the operators $\Pi_n$ from the
experimentally sampled probability distribution $p_n^\rho$, over a
suitable set of known states $\rho$.
These must form a tomographically complete set, spanning the 
Hilbert subspace where the POVM elements are 
defined on \cite{DAriano2000}. 
\par
The simplest
choice for a continuous variable system is a set of coherent states. They
provide an overcomplete basis for the Fock space, and it has already
been proved that even 1-dimensional discrete collections of coherent
states form a complete basis, and may be used to reconstruct classical
and non-classical states \cite{Janszky1990, Janszky1995}.
In fact, the experimental distributions of the outcomes 
for a set of coherent states already provide a full representation of
the detector operators, in the form of a sample of their Q-functions
$$ Q_n (\alpha) = \frac{1}{\pi} \bra{\alpha} \Pi_n \ket{\alpha} =
\frac{1}{\pi} \mathcal{P}_n^\alpha\,,$$ where
$\{{\mathcal{P}^\alpha_n}\}$
represents the probability distribution for 
a coherent state. However, this 
representation is not suitable to provide a complete and 
reliable {\em characterization} of the detector. In fact, 
any subsequent use of this reconstruction scheme to {\em predict} the
outcome of the measurement for a different signal would involve the
(numerical) evaluation of the trace rule in the phase-space as 
$$ \tr{\rho\, \Pi_n} = \int_{\mathbb C}\! d^2 \alpha\, P[\rho] (\alpha)\, Q_n
(\alpha)\,,$$
where the Glauber P-function $P[\rho] (\alpha)$ is singular for any
nonclassical state, and thus not suitable for sampling.
In order to overcome this problem we suggest an expansion in the quadrature basis of the operator measure 
associated with the detector, using as probe  
a set of coherent states. We then obtain the coefficients of the 
expansion using a least squares
algorithm on a sufficiently large sample of data. 
We also validate the experimentally obtained POVM by
reconstructing nonclassical known states. 
Finally, we exploit results to estimate the
overall quantum efficiency of the detector.
\par
The paper is structured as follows. In Section \ref{s:hdm} we 
review the description of homodyne detection and introduce 
the algorithm employed for the reconstruction of its POVM. 
In Section \ref{s:exp} we describe our experimental apparatus, 
whereas in Section \ref{s:hdt} we present results of the 
reconstruction, as well as their validation on nonclassical states. 
Section \ref{s:out} closes the paper with some 
concluding remarks.
\section{Homodyne Detection}\label{s:hdm}
A homodyne detector is a fully phase
sensitive apparatus that provides a complete characterization of any
given state of a single-mode radiation field \cite{Leonhardt1997}. This
state, the {\it signal}, is sent to a balanced beam splitter, where it
interferes with an intense coherent field, the {\it local oscillator},
usually coming from the same laser source. The phase of the signal has
then a precise value $\phi$ with respect to the local oscillator, and
can be adjusted by means of a piezo-actuated mirror. The two outputs of
the beam splitter are then focused on two photodiodes, and the resulting
photocurrents subtracted and analyzed. It can be shown that, in the
approximation of high amplitude $\abs{\beta}$ of the local oscillator, the
measurement
associated to this detector corresponds to
\begin{equation}
\label{eq:hom}
{X} = \frac{{a} {b}^\dagger + {a}^\dagger 
{b}}{\sqrt{2} \left | \beta \right |} \underset{\left | \beta 
\right | \gg 1} {\longrightarrow} \frac{{a}^\dagger 
e^{\imath \phi} + {a}\, e^{-\imath \phi}}{2} = {x}_\phi
\end{equation}
where ${a}$ and ${b}$ are the mode operators for
the signal and the local oscillator, respectively. The operator
${b}$ was replaced by $\abs{\beta}e^{\imath \phi}$ in Eq.~(\ref{eq:hom})
 by considering its action onto the local oscillator, that
can be treated as a coherent state $\ket{\beta}$. The operation connected
to the working scheme of this detector is therefore the measurement of
the quadrature operator ${x}_\phi$ on the signal mode. Such a link
states the equivalence between the discrete spectrum of the operator
${X}$ and the continuous one of the quadrature, due to the high
intensity of the local oscillator, that can be 
consequently treated classically. This equivalence can be extended 
to the characteristic functions
\begin{equation}
\Tr{a,b}{e^{ \I \lambda {X}} \rho \otimes \proj{\beta}{\beta} } 
\underset{\left | \beta \right | \gg 1} {\longrightarrow} 
\Tr{a}{e^{ \I \lambda {x}_\phi} \rho }
\end{equation}
assuring the equivalence of all moments \cite{Paris2004}.
\par
The key feature of homodyne detection is its ability to
discern between different phase values of an input signal, setting
itself apart from the photon counting detectors that have been
characterized in the past. A straightforward choice for a basis in which
representing the POVM elements of a phase insensitive device is the
number basis, in the form 
$\Pi_k = \sum_{n=0}^{+\infty} \pi_n ^{(k)} 
\proj{n}{n}$, where $\proj{n}{n}$ is the projector onto the n-photon Fock state.
Such a description is no longer suitable for our apparatus, that hinges
on a phase-sensitive operation scheme. Off-diagonal elements in the
number-basis expansion could enclose phase-sensitive properties, as was
done in \cite{Zhang2012}, but reconstruction of the detector operators
would become increasingly difficult due to the high dimension of the
Hilbert space the POVM would be defined in. 
\subsection{The reconstruction algorithm}
\begin{figure}[h!]
\includegraphics[width=1\columnwidth]{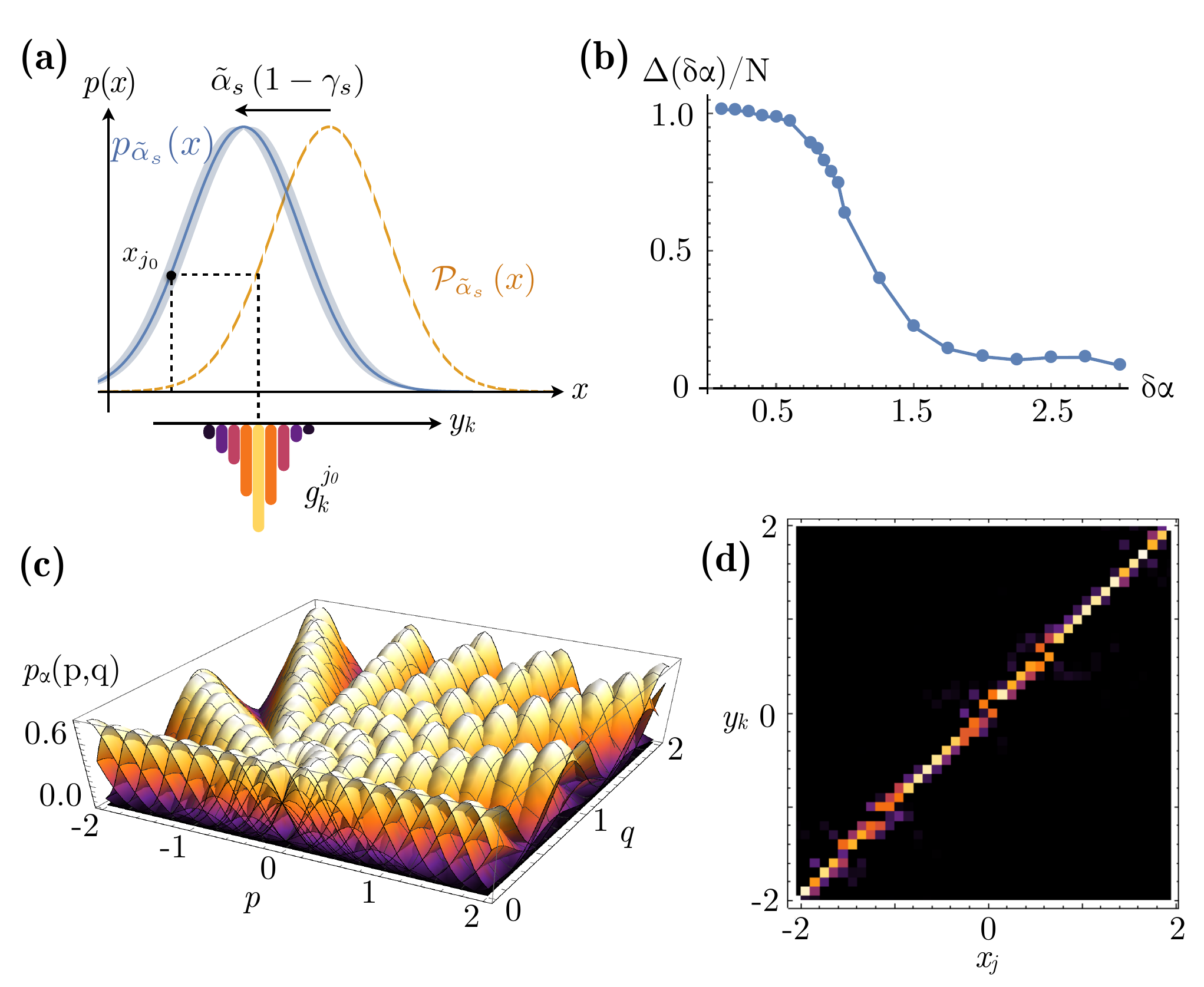}
\caption{(Color online) 
Quantum tomography of a homodyne detector. 
A pictorial representation of the least-squares
algorithm is reported in panel (a). The experimentally sampled
probability distributions $p_{\tilde{\alpha}} (x)$ (solid blue) are
compared to the Q~-~function representation
$\mathcal{P}_{\tilde{\alpha}} (x)$ (dashed yellow). This is shown 
dashed
to visualise the discretisation of Eq. (\ref{eq:model}). Panel (b)
shows the efficiency of reconstruction as a function of increasing amplitude
spacing in the tomographic set. The
function $\Delta(\delta \alpha)$ is here normalized to the matrix
dimension $N$. A sharp transition can be seen as the spacing between the
states of the
set approaches the maximum for the overlap function of Eq.
(\ref{eq:Overlap}). 
Panel (c) shows the phase space representation  
of the full set of coherent states used for tomography.
Panel (d) shows the result of the reconstruction obtained
imposing $\abs{\alpha_s}=\abs{\tilde{\alpha}_s}$. Notice the 
(expected) diagonal shape.}
\label{fig1}
\end{figure}

A suitable basis to expand the POVM $\Pi (x)$ 
of a phase-sensitive detector  for continuous 
variable systems is the set 
$\left \{ \proj{y}{y} \right \}_{y \in \mathbb{R}}$ 
of eigenstates of a quadrature operator 
${x}_\phi$, e.g. setting $\phi=0$
\begin{equation}
\label{eq:model}
\Pi (x) = \int \! dy\, g^{(x)}_y \proj{y}{y}\,.
\end{equation}
The detector has now a Q-function representation given by
\begin{align}
\label{eq:qrep}
\mathcal{P}_\alpha (x) &= \bra{\alpha} \Pi (x) 
\ket{\alpha} \nonumber \\ & = \int \! \!   
dy\, g^{(x)}_y \sqrt{\frac{2}{\pi}}\, 
\exp \left\{-2 ( y - \abs{\alpha}) ^2\right\}\,,
\end{align}
where the matrix $g^{(x)}_k$ contains all the information 
needed to describe the response of the detector.
The
set of operators $\left \{ \Pi (x) \right \}$ can then be discretized,
reflecting the experimental sampling during the measurement process, and
be confined to a selected portion of the quadrature range, say 
$\left [ x_{\min}, \,
x_{\max} \right ]$. The expansion on the quadrature basis $\left \{
\proj{y}{y} \right \}$ can be discretized as well,
reducing the number of POVM elements.
Equation (\ref{eq:qrep}) may be rewritten as
\begin{equation}
\label{eq:modeld}
\Pi_j \equiv \Pi(x_j) = \sum_k g^{j}_k \proj{y_k}{y_k}\,.
\end{equation}
The analogue of 
Eq. (\ref{eq:qrep}), i.e. ${\mathcal P}_\alpha (x_j) = \langle \alpha|
\Pi (x_j)|\alpha\rangle$,
can now be compared to experimental results and 
may be inverted to find the matrix
$\mathbf{g}$.  To this purpose, starting from a set of coherent states
with calibrated amplitudes $\left \{ \abs{\tilde{\alpha}_s} \right \}$,
we use a least-squares method:
\begin{equation}
\label{eq:algo}
g^{j}_k = \underset{ \left \{ g_k^j \right \} >0 }{
\mathtt{arg\, \, min}} \left \{ \sum_j \sum_s 
\left [ p_{\tilde{\alpha}_s} (x_j) - 
\mathcal{P}_{\tilde{\alpha}_s } (x_j) \right ] ^2 \right \} \,,
\end{equation}
where $p_{\tilde{\alpha}_s} (x_j)$ is the experimentally observed 
distribution for the coherent state 
with amplitude $\abs{\tilde{\alpha}_s}$.
A pictorial representation of the algorithm is presented in Fig.
\ref{fig1}(a).  The algorithm retrieves the matrix
$\mathbf{g}$ by comparing the experimentally sampled 
distributions to the corresponding 
Q~-~function representation.
Starting from a coherent state of amplitude $\abs{\tilde{\alpha}_s}$ and
a quadrature value $x_j$, the algorithm finds the set of
coefficients $g^j_k$ that minimises the difference between
$p_{\tilde{\alpha}_s} (x_{j})$ and $\mathcal{P}_{\tilde{\alpha}_s}
(x_{j})$. Experimental imperfections and fluctuations, 
corresponding to the shaded blue area in Fig. \ref{fig1}(a), 
degrade this relationship and ``switch on''
new coefficients in the expansion of Eq. (\ref{eq:qrep}). 
At the same time, each calibrated amplitude $\abs{\tilde{\alpha}_s}$ is
associated to a rescaled value $\abs{\alpha_s}=\abs{\tilde{\alpha}_s}
\gamma_s$, 
to explicitly include the detector finite efficiency.
The least squares algorithm of Eq.
(\ref{eq:algo}) performs this minimisation simultaneously for all the
quadrature values $x_j \in \left [ x_{\min}, \, x_{\max} \right ]$ and all
the coherent states in the set. The detector response is then fully
contained in the matrix $\mathbf{g}$.
\par
If we look back at Eq. (\ref{eq:model}), it is quite natural to link
the characteristics of the matrix ${\mathbf g}$ to the features of 
the detector reconstruction.  Each matrix row $g^j_k$
associates a quadrature value $x_{j}$ to a set of projectors on
quadrature eigenstates, with weights given by the coefficients in Eq.
(\ref{eq:modeld}). For an ideal detector, the matrix is diagonal
$g^j_k=\delta_{kj}$, i.e. the only nonzero coefficient
associates a quadrature value $x_j$ to its projector $\proj{x_j}{x_j}$.
As we have seen, experimental imperfections will degrade this one-to-one
relationship, spreading the coefficients around a central value.  At the
same time, the full array of coefficients $\gamma_s$ provides a unified
model for the response of the detector.  A detector tomography
devised in this way 
is general enough for application to different configurations of
the homodyne detection.
\par
Let us now focus on the tomographic set, and in particular on the 
characteristics required to perform a reliable reconstruction. To 
this aim we have performed simulated 
experiments with sets having an increasing number of equidistant 
coherent states, with
amplitudes in a given range $\alpha \in [-3,\,3]$, and measured at the same
phase $\phi=0$. For each set the matrix representation of the detector
$g^{j}_k$ was retrieved, and associated to the amplitude spacing 
$\delta \alpha$ between the coherent states in the set. Since the 
identity matrix ${\mathbb I}$ is the
ideal-case solution of the reconstruction algorithm,
we consider the following function of the 
amplitude spacing $\delta \alpha$
\begin{equation}
\label{eq:MinRef}
\Delta (\delta \alpha) = \tr{ \mathbb{I} \cdot 
\left [ g^{j}_k \right ]_{{\delta \alpha}}}
\end{equation}
as a figure of merit to assess tomographic sets, e.g. to 
find the minimal $\delta\alpha$ corresponding to 
a reliable reconstruction. 
\par
For a perfect reconstruction, the two matrices are
both the identity and $\Delta$ reduces to the dimension of the matrix
$\mathbf{g}$. In the opposite case, more and more elements on the
diagonal will be voided, and the trace will decrease. In Fig.
\ref{fig1}(b) we show the results: a steep transition, corresponding to
a deterioration of the reconstruction, appears for $\delta \alpha \simeq
0.7$. A similar conclusion may be obtained theoretically upon
considering the overlap of two Gaussian $p_\sigma(x,x_0)$ of 
equal standard deviation $\sigma$ but
varying center $x_0$, i.e. 
\begin{equation}
\label{eq:Overlap}
f(y) = \sqrt{\pi } \int_{-\infty }^{+\infty } 
\!\!\!\!\!\!\!dx\, p_\sigma(x,x_0) \, p_\sigma(x,x_0 + y)\,.
\end{equation}
In particular, the function $y \cdot f(y)$ may be used to assess 
the tomographic set of coherent states, as it captures, roughly speaking, 
the trade-off between an increasing spacing and a
decreasing overlap. Upon substituting $\sigma=0.5$, as it is 
for coherent states, we have that $y \cdot f \left (
y \right )$ has a maximum at $y_{\max} = 2^{-1/2} 
\simeq 0.707$, in good agreement
with the value obtained by simulated experiments via 
Eq. (\ref{eq:MinRef}). The reliability of this estimation has been 
then confirmed experimentally (see below).
\section{Experimental apparatus}\label{s:exp}
\begin{figure}[h!]
\includegraphics[width=1\columnwidth]{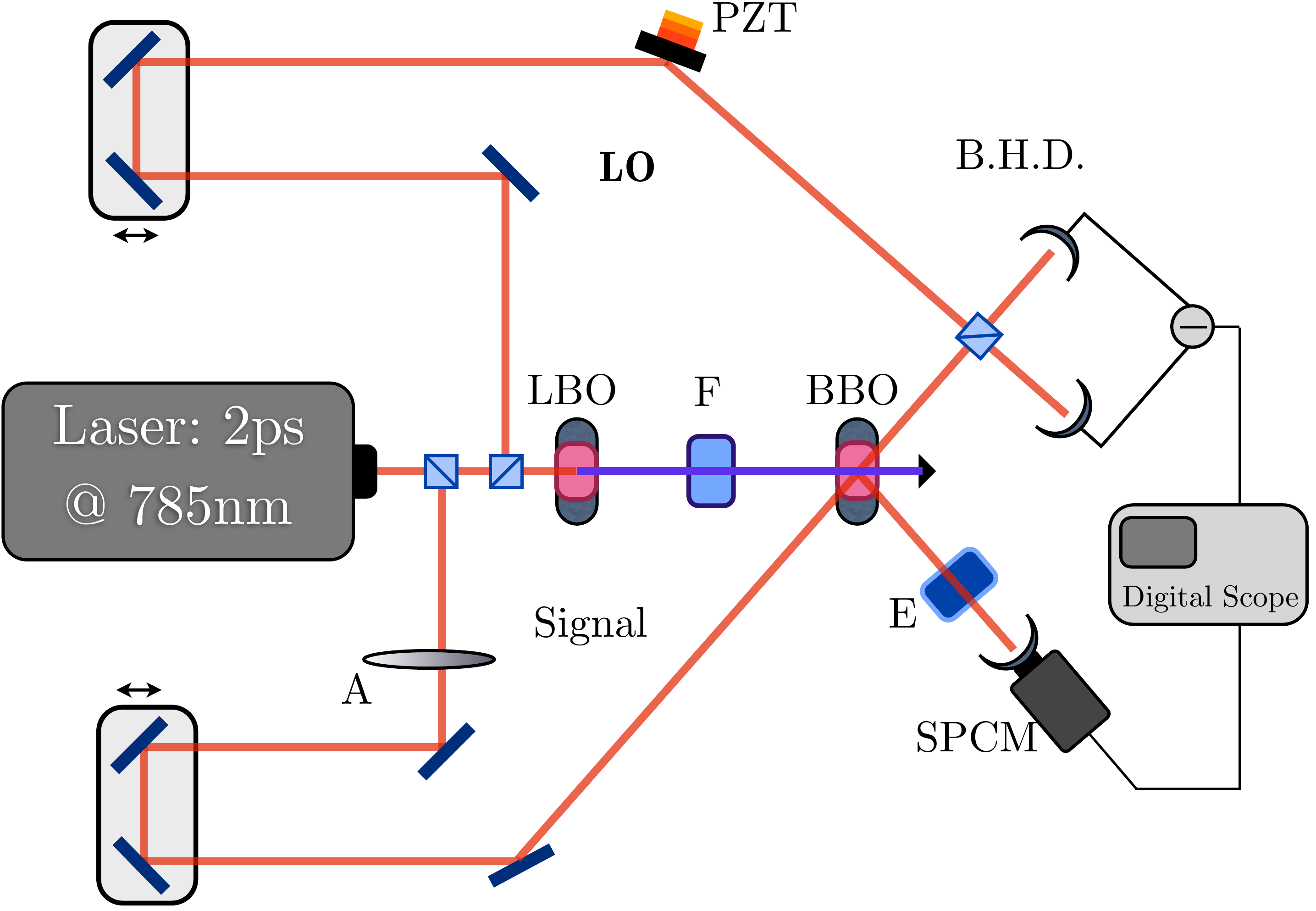}
\caption{(Color online) Schematic diagram of the experimental setup.
Picosecond-duration pulses at 785 nm are split
by two high-transmission beam splitters. A first beam serves as the
local oscillator (LO) for balanced homodyne detection (B.H.D.), while a
second provides the probe coherent states for detector tomography.
Optical delay lines provide synchronization, while the probe amplitude
is adjusted with the attenuator (A). The main portion of the beam is
frequency doubled in a LBO crystal, and then spatially and spectrally
filtered by means of a pin-hole and a Michelson interferometer (F). It
finally serves as the pump for a type-I BBO crystal for parametric down
conversion, whose idler mode is used for calibration means and, after
narrow Etalon spectral filtering (E), as a trigger signal.}
\label{fig2}
\end{figure}

A schematic diagram of the experimental setup is shown in 
Fig. \ref{fig2}. The apparatus is based on a
mode-locked Ti:sapphire laser (Spectra-Physics Tsunami) providing,
after suitable splitting, both the local oscillator 
(LO) beam for balanced homodyne
detection and the probe coherent states
for detector tomography. 
The laser emits $2$ ps
pulses at a central wavelength of $78$5 nm, with a repetition rate of
$82$ MHz. 
The detector characterized in this paper is an optical homodyne
apparatus, operating in the time domain at high sampling frequency
\cite{Zavatta2002,Zavatta2006}. Amplitudes of the probe coherent states
were selected by means of a reflective-coating glass attenuator. 
Precise
calibration of each state is done by means of a Type I BBO crystal cut
for degenerate spontaneous down-conversion (SPDC), pumped by the frequency
doubled portion of the main laser beam. 
The injection of the probe coherent states
into the signal path of the SPDC triggers the stimulated
emission of downconverted photon pairs in the same mode
(thus generating single-photon-added coherent states, SPACSs
\cite{Zavatta2004}) and in the idler mode, 
at a rate proportional to 
$1+\abss{\alpha}$, where $\alpha$ is the amplitude of the
incoming coherent state. Such a procedure provides a precise
standard-free calibration of the input amplitude \cite{Migdall1999} 
by means of the ratio between the count rate of
stimulated and spontaneous events
pursuing the idea of a calibration-free characterization.
\par
\section{Detector Tomography}\label{s:hdt}
In our detector tomography 
we have focused attention to the range 
$x \in \left [ -2, \, 2 \right ]$ probed by a
 tomographic set of coherent states with amplitudes $\alpha_s \in \{
 -3,\,3 \}$. Our set is made of coherent states with $12$ different 
 amplitudes, each one measured at $9$ different phase values 
 between 0 and $\pi$. The full set is represented in 
 Fig.~\ref{fig1}(c). 
\par 
 As a first step we have performed a preliminary validating step 
 of our reconstructing algorithm, by neglecting the rescaling factors
 $\gamma_s$. The set of amplitudes $\left \{ \alpha_s \right \}$
 has been measured with the homodyne detector, as the mean 
 value of the probability distributions, and the coefficient 
 matrix $\mathbf{g}$ has been retrieved imposing 
 $\abs{\tilde{\alpha}_s}=\abs{\alpha_s}$. The result, 
 reported in Fig.~\ref{fig1}(d), presents the
 expected diagonal shape, partially blurred by fluctuations. The homodyne
 detection model obtained in this way may efficiently describe 
 the detector behavior on several classical and non classical states. 
\par
\begin{figure}[h!]
\centering
\includegraphics[width=1\columnwidth]{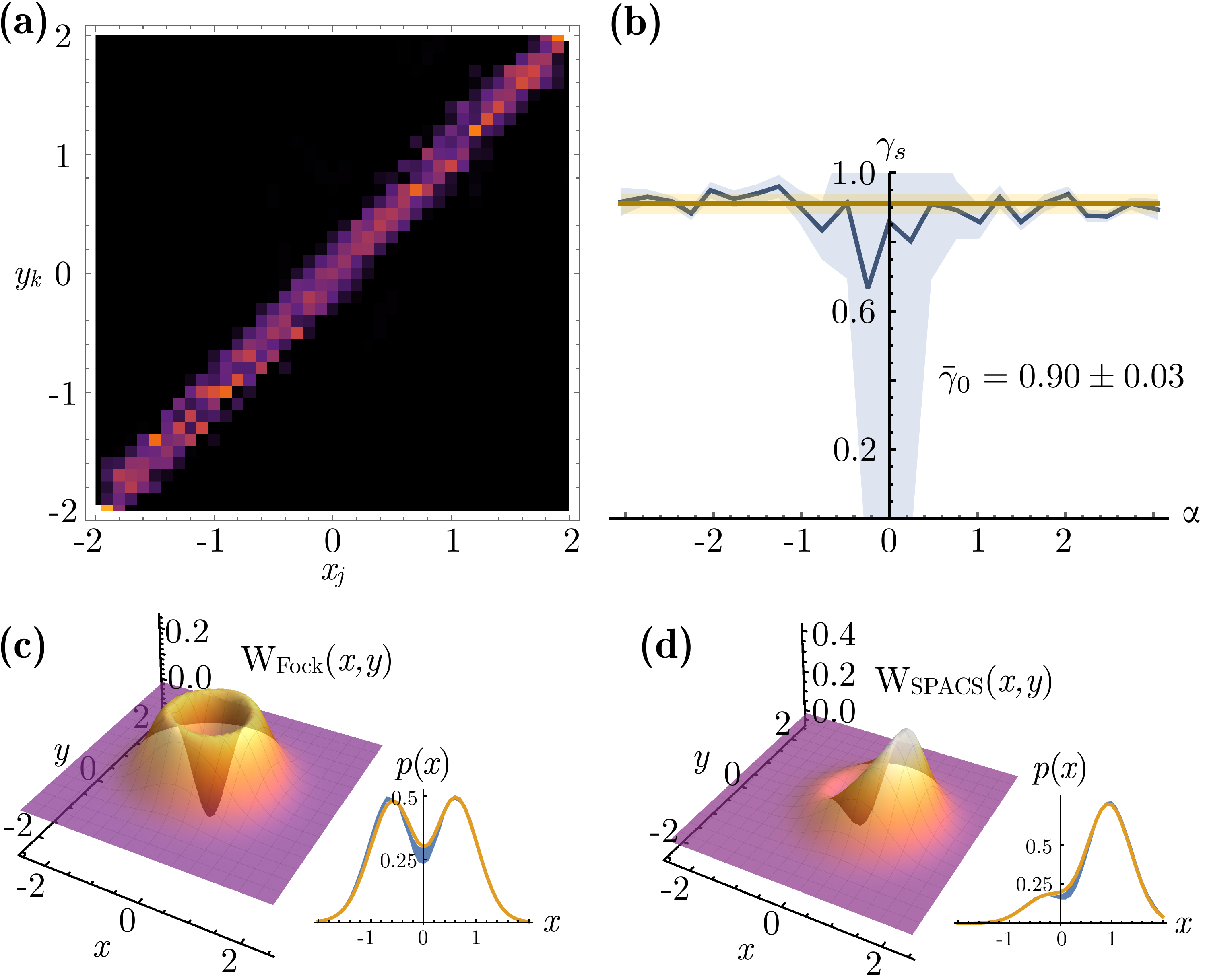}
\caption{(Color online) 
Quantum tomography of a homodyne detector. 
Panel (a) shows the matrix $g_k^{j}$ representing the expansion 
of the POVM elements in the quadrature basis. By comparison of this 
matrix with that of Fig. \ref{fig1}(d) one notices the spread of the
coefficients around the diagonal, due to the additional parameters
$\gamma_s$. These are reported in blue in the plot of Fig.
\ref{fig3}(b), where the shaded area represents an estimate of the
uncertainty, obtained 
through the second derivative of the function in Eq. (\ref{eq:algo}).
In a similar fashion, the mean value $\bar{\gamma}_0$ and its 
uncertainty are indicated in solid and shaded yellow. 
These results were then validated on the reconstruction of 
nonclassical states. In panels (c) and (d) we report the 
experimentally reconstructed Wigner
functions for the Fock state $|1\rangle$ and for a single photon-added 
coherent state. The side plots in both panels
show the Wigner
marginal distributions (yellow), which are both very close to their
representations in the detector description (the uncertainty 
due to the spread of $\bar{\gamma}_0$ is given by the
blue-shadowed areas).}
\label{fig3}
\end{figure}

Upon
 confirmation of the effectiveness of our method, we selected a reduced
 collection of coherent states, concentrating on those with a null or
 $\pi$ phase difference between the signal and the local oscillator.
 This set of $25$ coherent states has been then inserted into the
 algorithm of Eq.~(\ref{eq:algo}) together with the set of calibrated
 amplitudes $\left \{\tilde \alpha_s \right \}$, and the results are
 shown in Fig.~\ref{fig3}(a). The
 retrieved matrix showed the expected diagonal shape, with an
 experimental spread around the central value. The array of coefficients
 $\gamma_s$ is reported in Fig.~\ref{fig3}(b).
 An estimate for the uncertainty is obtained from the inverse of the 
 second derivative of the function minimized in the least-squares algorithm.
This uncertainty is increasing for decreasing values of $\alpha$,
since for the limiting case of $\alpha=0$ our model is not
defined. It was then used to weight the $\gamma_s$, and get a final mean
value of $\bar{\gamma}_0=0.90 \pm 0.03$. 
\par
\subsection{POVM validation}
The aim of quantum detector tomography is to fully characterize 
a given detector by retrieving the set of operators that fully
describes its measurement process. In order to validate our technique 
for QDT, we have employed the reconstructed POVM to reproduce
measurements performed on known states. In particular, we have tested 
the POVM on two kinds of nonclassical states: a single-photon Fock state
and a SPACS with $\alpha=0.5$. These two states have been 
first measured in our
setup, and the experimental results have been then compared to those
obtained for the same states using the reconstructed detector POVM. 
The photon addition scheme of
our setup, fundamental for the generation of these
two states, had been previously characterised \cite{Zavatta2004a}, and
the operator $\hta$ had been found to apply 
to a given input state with preparation
efficiency $\zeta \approx 0.91$. The measured quadrature 
distributions for the Fock state and the SPACS, 
reported in Fig. \ref{fig3}(c) and \ref{fig3}(d), 
are in excellent agreement with the
expected Q-function representation based on the tomography of our
homodyne detector.
The robustness and reliability of our method has been 
thus confirmed and 
we proved that the specific
experimental realization of the detector, which depends
on several parameters (like the detrector quantum efficiency, the degree
of mode matching, the alignment, etc.), can be efficiently captured by
the tomographic procedure. We also proved that the results of subsequent
measurements can be effectively reproduced.
\par
\begin{figure}[h!]
\centering
\includegraphics[width=1\columnwidth]{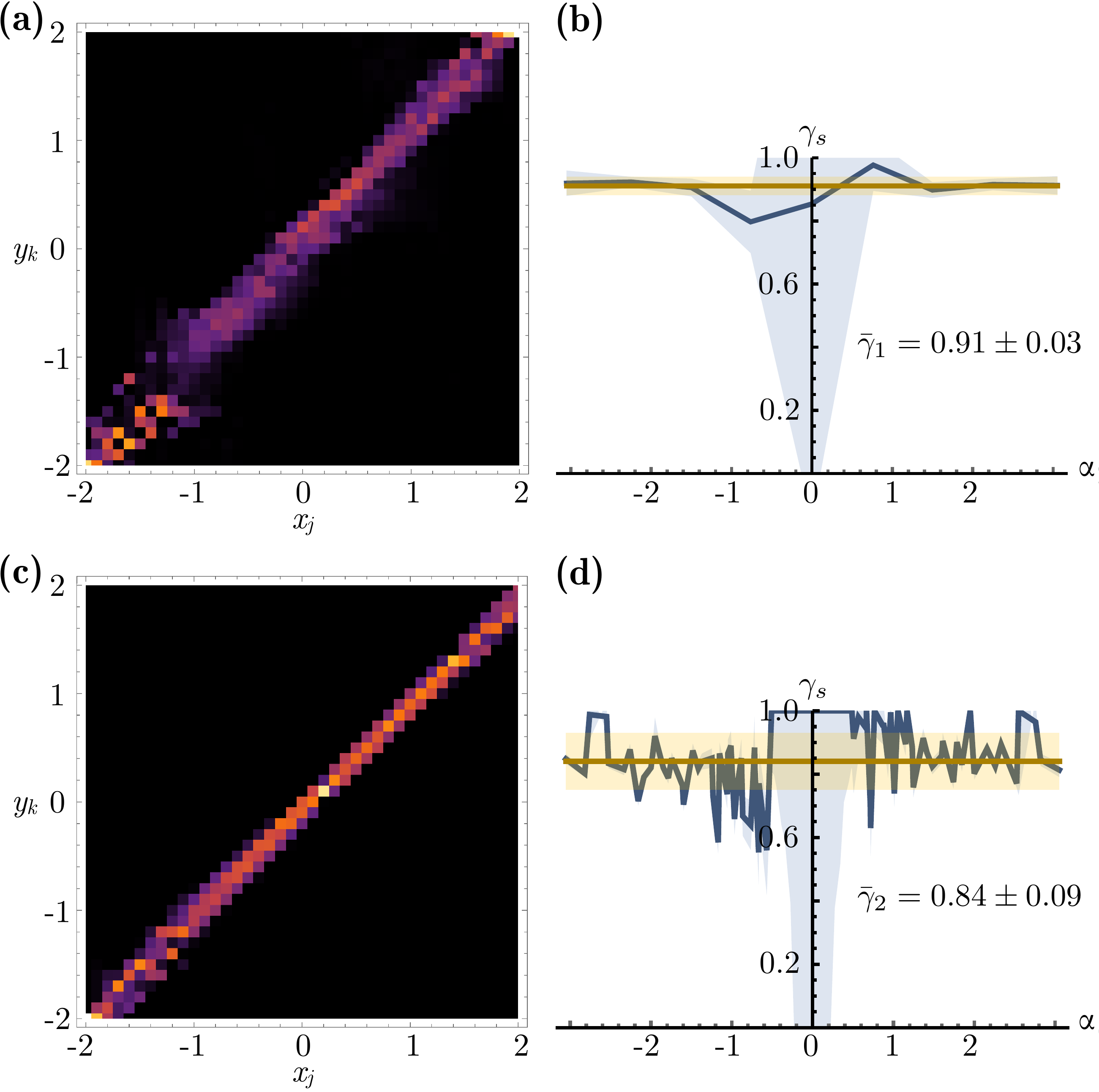}
\caption{(Color online) Changing the number of coherent states in the
tomographic set. Panels (a) and (b) show results of a
successful tomography for the limiting case of an amplitude spacing of
$0.7$, as predicted by the theoretical modeling. The limiting case of
oversampling was also tested, yielding another positive result. 
Results of QDT using the full tomographic set of $109$ states are 
reported in panels (c) and (d).} 
\label{fig4}
\end{figure}
\par
The tomographic set of coherent states that we have used throughout the
paper has proved to be a good test for our detectors. In order to
improve the accuracy of the reconstruction we may extend the set to
cover a bigger portion of phase space, while to minimize the
experimental effort we may want to reduce the number of states in the
set. Proceeding as above, and following our predictions from Eq.
(\ref{eq:MinRef}), we found that a minimal set of nine coherent states may
be selected from the experimental data, all at phase zero.  The
amplitude spacing is three times larger than in the previous situation,
but the reduced set is still able to provide a quorum for the
tomography. Results are presented in Fig. \ref{fig4}(a) and (b). On the
other hand, even with the full set of $109$ coherent states we have been 
able
to efficiently reconstruct the detector, despite the increased phase space
coverage and the increased fluctuations, due to oversampling. The
matrix $g^{j}_k$ and the coefficients $\gamma_s$ for this case 
are shown in Fig. \ref{fig4}(c) and (d).
\par
On the basis of the previous analysis and considering the 
large uncertainty of data in the area close to the origin, 
we have assigned a fixed value $\gamma_s=1$ to probe
states with amplitudes smaller than 0.5. We found that the values of 
$\bar{\gamma}$ for the smallest and the largest tomographic
set are given by $\bar{\gamma}_1=0.91 \pm 0.03$ and
$\bar{\gamma}_2=0.84 \pm 0.09$ respectively. In fact, all the values of 
$\bar{\gamma}_0$, $\bar{\gamma}_1$ and $\bar{\gamma}_2$ are comparable 
within the uncertainty. On the other hand, they
convey different information regarding the detector. The increased
coefficient spread of the matrices reported in Fig. \ref{fig3}(a) and Fig. 
\ref{fig4}(a) can
be considered as an additional rescaling parameter, modeling the
experimental fluctuations, that is therefore directly included in the
tomography. Values for $\bar{\gamma}$ are then larger, with reduced
uncertainty. The matrix of Fig. \ref{fig4}(c) has instead a smaller
spread, and therefore the extra rescaling is conveyed in
$\bar{\gamma}_2$, lowering its value and making it more accurate, even
though less precise.  Thus $\bar{\gamma}_2$ is better suited to be
compared to the value of the quantum detection efficiency 
that can be obtained by classical calibration. Indeed, we find that
$\gamma_2 \approx \sqrt \eta$, despite the fact that our model does not
involve any prior knowledge of the detector structure or implementation.
Different sets can therefore be used to highlight
specific properties of the detector, adding value to our technique.
\section{Conclusions}\label{s:out}
We have suggested and demonstrated a full quantum detector tomography
technique for a homodyne detector. In ideal conditions each detector
operator is associated to a single quadrature projector: our technique
suitably describes how experimental noise and specific physical
realizations of the detector affect this description and allows us to
quantify experimentally the spreading of the detector operators onto
adjacent quadrature states. The model is general enough to describe any
kind of homodyne setup, and it has proven capable of effectively
describing the detector response to different tomographic sets. 
The reconstructed POVM have been then validated on different 
nonclassical states, thus confirming the robustness and the 
reliability of the method.
\par
Our results provide a general method to estimate the overall
detection efficiency in this class of detectors and may represent a
valuable resource to optimize homodyne detection in different
situations. Our model may be generalized to specifically treat 
single parameters of homodyne detectors, as mode mismatch or 
correlations between amplitude and phase noise.
Besides, a better understanding of the fundamental functioning 
of this detector paves the way to an evolution of the same, 
as well as a broader and more precise use in quantum optics 
and quantum technology
with continuous variables. 
\section*{Acknowledgments}
This work has been supported by UniMI through the H2020 Transition Grant
15-6-3008000-625 and by EU through the Collaborative Project H2020 QuProCS
(Grant Agreement 641277).
MB and AZ acknowledge the support of Ente Cassa di Risparmio di Firenze
and of the Italian Ministry of Education, University and Research
(MIUR), under the 'Progetto Premiale: Oltre i limiti classici di misura'
\bibliography{QDTHDlib.bib}
\end{document}